%% file: mainvldb.tex
\documentclass[twoside,10pt]{article} 
\usepackage{amsfonts}
\usepackage{amsmath}
\usepackage{amssymb}
\usepackage{epsfig}
\usepackage{graphicx}
\usepackage{latexsym}
\usepackage{subfigure}
\usepackage{times}
\usepackage{url}
\usepackage[usenames, dvipsnames]{color}
\usepackage{xcolor}
\usepackage[colorlinks=true,citecolor=blue,linkcolor=blue,urlcolor=blue]{hyperref}

\newcommand{\app}{\textsc{Seagull}}

\makeatletter
\providecommand*{\cupdot}{%
  \mathbin{%
    \mathpalette\@cupdot{}%
  }%
}
\newcommand*{\@cupdot}[2]{%
  \ooalign{%
    $\m@th#1\cup$\cr
    \hidewidth$\m@th#1\cdot$\hidewidth
  }%
}

\usepackage[ruled]{algorithm}
\PassOptionsToPackage{noend}{algpseudocode}
\usepackage{algpseudocode}

\renewcommand{\algorithmiccomment}[1]{\bgroup\hfill//~#1\egroup}

\algnewcommand\algorithmicswitch{\textbf{switch}}
\algnewcommand\algorithmiccase{\textbf{case}}
\algnewcommand\algorithmicassert{\texttt{assert}}
\algnewcommand\Assert[1]{\State \algorithmicassert(#1)}%
\algdef{SE}[SWITCH]{Switch}{EndSwitch}[1]{\algorithmicswitch\ #1\ \algorithmicdo}{\algorithmicend\ \algorithmicswitch}%
\algdef{SE}[CASE]{Case}{EndCase}[1]{\algorithmiccase\ #1}{\algorithmicend\ \algorithmiccase}%
\algtext*{EndSwitch}%
\algtext*{EndCase}%

\usepackage[font=small,labelfont=bf,tableposition=top]{caption}

\usepackage{listings}
\renewcommand{\algorithmiccomment}[1]{/* #1 */}

\usepackage{amsthm}
\usepackage{setspace,xspace}
\graphicspath{{figures/}}

\newcommand{\nop}[1]{}

\newcommand{\rem}[1]{\marginpar{\flushleft{#1}}}
\renewcommand{\rem}[1]{} 

\renewcommand{\algorithmiccomment}[1]{/* #1 */}

\graphicspath{{figures/}}
\newtheorem{definition}{Definition}

\usepackage{multirow}
\newcommand{\eat}[1] {}

\usepackage{fancyhdr}

\fancypagestyle{plain}{%
  \fancyhead{}
  
}

 \newlength{\hoehe}
 \newlength{\breite}
\title{\fontsize{15}{15}\selectfont Seagull: An Infrastructure for Load Prediction and\\ Optimized Resource Allocation\\\vspace*{1cm}
\large Technical Report\\
August, 2020\\
\vspace*{1cm}
{\large 
Olga Poppe, 
Tayo Amuneke, 
Dalitso Banda,
Aritra De,
Ari Green, 
Manon Knoertzer,
Ehi Nosakhare, 
Karthik Rajendran, 
Deepak Shankargouda,
Meina Wang, 
Alan Au,
Carlo Curino,
Qun Guo,
Alekh Jindal,
Ajay Kalhan,
Morgan Oslake,
Sonia Parchani,
Vijay Ramani, 
Raj Sellappan, 
Saikat Sen,
Sheetal Shrotri,
Soundararajan Srinivasan,
Ping Xia,
Shize Xu,
Alicia Yang, and
Yiwen Zhu
}}
\date{\Large 
\large Microsoft Corporation, One Microsoft Way, Redmond, WA 98052\\
firstname.lastname@microsoft.com
\vfill
}

\begin{document}
\maketitle

\begin{spacing}{0.8}
{\footnotesize \noindent \textbf{Copyright} \copyright{} 2020 by
Microsoft. Permission to make digital or hard copies of all or
part of this work for personal use is granted without fee provided
that copies bear this notice and the full citation on the first
page. To copy otherwise, to republish, to post on servers or to
redistribute to lists, requires prior specific permission. }
\end{spacing}

\clearpage
\pagestyle{fancy}

\clearpage
\tableofcontents

\pagenumbering{arabic}
\setcounter{page}{1}

%
%

\newpage
\input{sections/abstract}
\input{sections/introduction}

\input{sections/infrastructure}

\input{sections/data}

\input{sections/accuracy}

\input{sections/ml}

\input{sections/evaluation}

\input{sections/lessons}
\input{sections/related_work}

\input{sections/conclusions}

\section*{Acknowledgements}

The authors thank Akshaya Annavajhala, Purnesh Dixit, Larry Franks, Chris Lauren, and Santhosh Pillai for fruitful discussions about AML. We are grateful to Ashvin Agrawal and Hiren Patel for their help with large-scale telemetry analysis. We thank Matteo Interlandi, Siqi Liu, and Markus Weimer for their feedback.

\bibliographystyle{abbrv}
\bibliography{mainvldb} 

\renewcommand{\thesection}{\Alph{section}}
\setcounter{subsection}{0}
\appendix
\input{sections/appendix}

\end{document}

%% file: sections/abstract.tex
\begin{abstract}

Microsoft Azure is dedicated to guarantee high quality of service to its customers, in particular, during periods of high customer activity, while controlling cost. We employ a Data Science (DS) driven solution to predict user load and leverage these predictions to optimize resource allocation. To this end, we built the \app\ infrastructure that processes per-server telemetry, validates the data, trains and deploys ML models. The models are used to predict customer load per server (24h into the future), and optimize service operations. \app\ continually re-evaluates accuracy of predictions, fallback to previously known good models and triggers alerts as appropriate. We deployed this infrastructure in production for PostgreSQL and MySQL servers across all Azure regions, and applied it to the problem of scheduling server backups during low-load time. This minimizes interference with user-induced load and improves customer experience. 

\end{abstract}

%% file: sections/introduction.tex
\section{Introduction}
\label{sec:introduction}


Microsoft Azure, Google Cloud Platform, Amazon Web Services, and Rackspace Cloud Servers are the leading cloud service providers that  aim to guarantee high quality of service to their customers, while controlling operating costs~\cite{cloudcomp, VCLRPKDH17}. Achieving these conflicting goals manually is labor-intensive, time-consuming, error-prone, neither scalable, nor durable.
Thus, these providers shift towards automatically managed services. 
To this end, Data Science (DS) techniques are deployed to predict resource demand and leverage these predictions to optimize resource allocation~\cite{RC}.

\textbf{Motivation}.
Backups of databases are currently scheduled by an automated workflow that does not take typical customer activity patterns into account. Thus, backups often collide with peaks of customer activity resulting in inevitable competition for resources and poor quality of service during backup windows. 
%
%
To solve this problem currently, an engineer plots the customer load per database per week and manually sets the backup window during low customer activity. However, this solution is neither scalable to millions of customers, nor durable since customer activity varies over time. More recently, customers can select a backup window themselves. However, they may not know the best time to run a backup. Instead of these manual solutions, DS techniques could be deployed to predict customer load. These predictions could then be leveraged to schedule backups during expected low customer activity. 

An infrastructure that analyses historical load per system component, predicts its future load, and leverages these predictions to optimize resource allocation is valuable for many products and use cases. 
Over the last two years we have built such an infrastructure, called \app, and applied it to two scenarios: (1)~Backup scheduling of PostgreSQL and MySQL servers and (2)~Preemptive  auto-scale of SQL databases (Appendix~\ref{sec:appendix}). These scenarios required us to battle-test the infrastructure across all Azure regions and gave us confidence on the high impact and generality of the \app\ approach.

\textbf{Challenges}. 
While building the \app\ infrastructure, we tackled the following open challenges.

$\bullet$ \textit{Design of an end-to-end infrastructure} that predicts resource utilization and leverages these predictions to optimize resource allocation. 
This infrastructure must be: (a)~Re\-usable for various products and application scenarios and (b)~Scalable to millions of customers worldwide.

$\bullet$ \textit{Implementation and deployment of this infrastructure to production} to predict customer activity and schedule backups such that they do not interfere with customer load.


$\bullet$ \textit{Accurate yet efficient customer low load prediction} for optimized backup scheduling. This challenge includes choice of an ML model that finds the middle ground between accuracy and scalability. In addition, prediction accuracy must be redefined to focus on predicting the lowest valley in customer CPU load that is long enough to fit a full backup of a server of its backup day. General load prediction per server per day is less critical for backup scheduling use case.


\textbf{State-of-the-Art Approaches}. 
While systems for ML were proposed in the past, most of them lack easy integration with Azure compute~\cite{mlflow, tensorflow, velox, clipper, caffe, mllib}. 
Thus, we built our solution using the functionality of Azure ML~\cite{aml}.


While time series forecast in general and load prediction in particular are well studied topics, none of the state-of-the-art approaches focused on predicting the lowest valley in customer CPU load for optimized backup scheduling. Instead, existing approaches focus on, for example, 
idle time detection for predictive resource provisioning~\cite{LRDXGKC16, VCLRPKDH17}, 
VM workload prediction for dynamic VM allocation~\cite{CMB14, RC}, and 
demand-driven auto-scale of resources~\cite{DLNK16,  quasar, dhalion, IKLL12, cloudscale, pstore, press}. 
Thus, these approaches do not tackle the unique challenges of low load prediction for optimized backup scheduling described above. In particular, they neither define the accuracy of low load prediction, nor compare several ML models with respect to low load prediction.

\begin{figure}[!t]
\centering
\includegraphics[width=0.7\columnwidth]{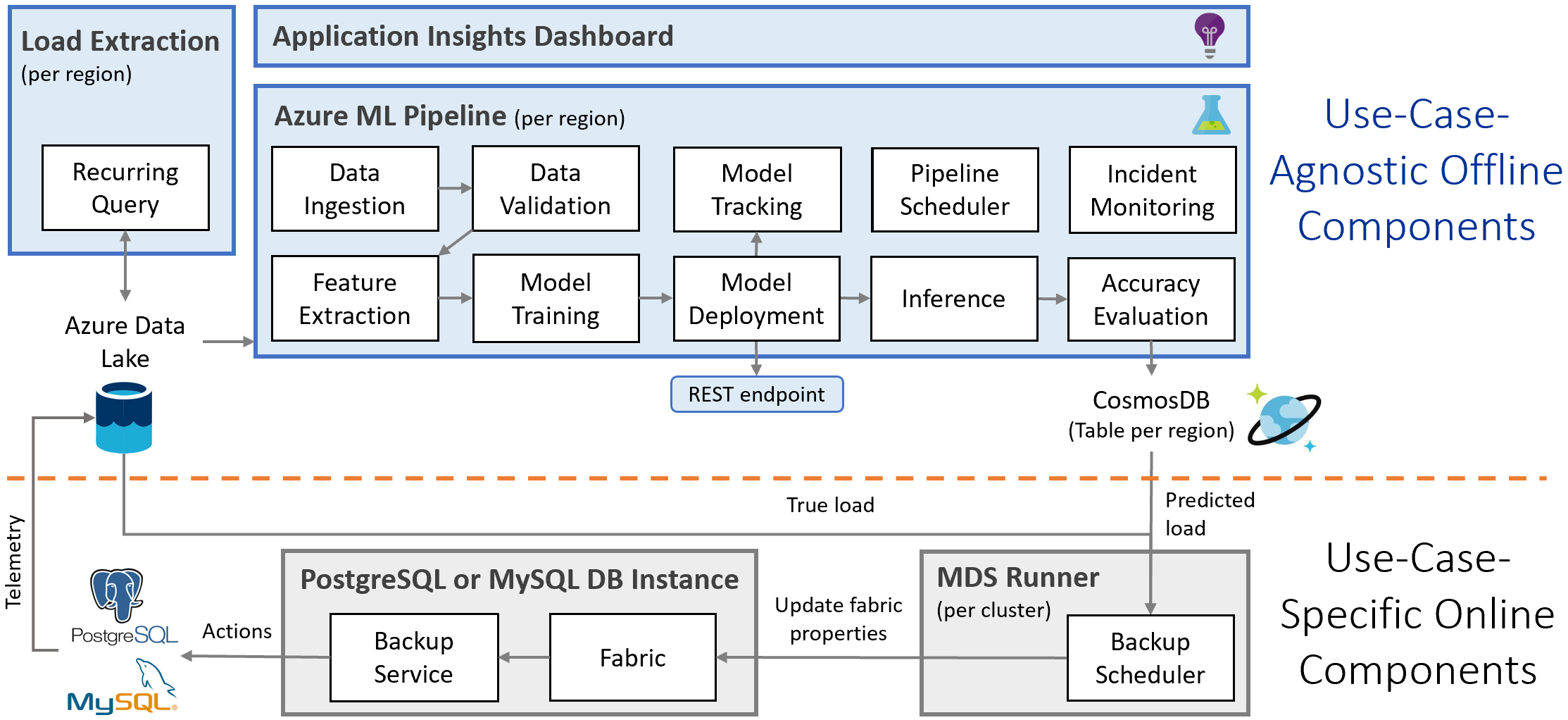}
\caption{\app\ infrastructure}
\label{fig:infrastrucute}
\end{figure}

\textbf{Proposed Solution}. 
We built the \app\ infrastructure (Figure~\ref{fig:infrastrucute}) that deploys DS techniques to predict resource utilization and leverages these predictions to optimize resource allocation. 
This infrastructure consumes prior load, validates this data, extracts features, trains an ML model, deploys this model to a REST endpoint, tracks the versions of all deployed models, predicts future load, and evaluates the accuracy of these predictions. 
%
%
We deployed \app\ to production worldwide to schedule backups of PostgreSQL and MySQL servers during time intervals of expected low customer activity.
We achieved several hundred hours of improved customer experience across all regions per month.


\textbf{Contributions}. 
Our \app\ approach features the following key innovations.

$\bullet$ We designed and  implemented an end-to-end \app\ infrastructure and deployed it in all Azure regions to optimize backup scheduling of PostgreSQL and MySQL servers. We describe our design principles and lessons learned. We describe our optimization techniques that reduce runtime and ensure scalability. We explain how to reuse the infrastructure for other scenarios. We evaluate the impact of the \app\ infrastructure on both improving customer experience and reducing engineering effort.



$\bullet$ We conducted comprehensive analysis to classify PostgreSQL and MySQL servers into homogeneous groups based on their typical customer activity patterns. Majority of servers are either stable or follow a daily or a weekly pattern. Thus, the load per server on previous (equivalent) day is a strong predictor of the load per server today. This heuristic, called persistent forecast, correctly predicted the lowest load window per server on its backup day in 96\% of cases. 


$\bullet$ We defined the accuracy of low load window prediction per server on its backup day as the combination of two metrics. One, the lowest load window is chosen correctly if there is no other window that is long enough to fit a full backup and has significantly lower average user CPU load. Two, the load during a lowest load window is predicted accurately if majority of predicted data points are within a tight acceptable error bound of their respective true data points.

$\bullet$ We applied several ML models commonly used for time series prediction (NimbusML~\cite{nimbusml}, GluonTS~\cite{gluonts}, and Prophet \cite{prophet}) to predict low load of unstable servers that do not follow a pattern that can be recognized by persistent forecast. We compared these models with respect to accuracy and scalability on real production data during one month in four Azure regions. Surprisingly, the accuracy of ML models is not significantly higher than the accuracy of persistent forecast. Thus, we deployed persistent forecast based on previous day to predict low load for all servers.



\textbf{Outline}.
We present the \app\ infrastructure in Section~\ref{sec:infrastructure} and classify the servers in Section~\ref{sec:load}.
Section~\ref{sec:accuracy} defines low load prediction accuracy, while
Section~\ref{sec:prediction} compares the ML models.
We evaluate \app\ in Sections~\ref{sec:evaluation},
summarize lessons learned in Section~\ref{sec:lessons} and,
review related work in Section~\ref{sec:related}.
Section~\ref{sec:conclusions} concludes the paper.

%% file: sections/infrastructure.tex
\section{Seagull Infrastructure}
\label{sec:infrastructure}

In this section, we summarize our design principles, give an overview of the \app\ infrastructure, and describe how to reuse it for other application scenarios.



\subsection{Design Principles}
\label{sec:principles}

\textbf{Modularity}. 
With the goal to reuse the \app\ infrastructure for various products and scenarios at Microsoft, we had to design it in a modular way. At the same time, we were determined to solve a specific task of optimized backup scheduling. To achieve both goals, we grouped the use-case-agnostic and use-case-specific components together (Figure~\ref{fig:infrastrucute}). The use-case-agnostic components can be reused in several scenarios (Section~\ref{sec:reuse}). For example, any ML model can be plugged in. Nevertheless, the use-case-agnostic components often have to be adjusted to a particular data set, product, and scenario. For example, if the load of most servers is stable or conforms to a business pattern (Section~\ref{sec:load}), a simple heuristic can be used to predict the load. Complex ML models may not be needed (Section~\ref{sec:prediction}). However, the usage patterns may change over time. This observation justifies the need for a robust infrastructure that automatically detects these changes, notifies about them, and allows to easily replace the model.

\textbf{Scalability}. 
With the goal to deploy the \app\ infrastructure in all Azure regions, we had to ensure that it scales well for production data. Thus, we broke the input data down by region and ran a DS pipeline per region. Since the size of regions varies, the size of input files ranges from hundreds of kilobytes to a few gigabytes. Consequently, the runtime of a pipeline ranges from few minutes to few hours (Figures~\ref{fig:training_inference_unstable} and \ref{fig:exp_runtime}). We used Dask~\cite{dask} to run time-consuming computations in parallel and achieved up to 4X speed-up compared to single-threaded execution (Figure~\ref{fig:runtime-dask}).  

The choice of an ML model is determined not only by its accuracy but also by its scalability. For example, ARIMA~\cite{arima} is computationally intensive since it searches the optimal values of six parameters per server in order to make an accurate load prediction per server. We had explored parameter sharing between servers but that resulted in a worsening of accuracy. While inference time is within a few seconds per server, fitting may take up to 3 hours per server. Hence, executing ARIMA in parallel for each server does not make runtime of ARIMA comparable to other models (Figure~\ref{fig:training_inference_unstable}). Thus, we excluded ARIMA from further consideration.

\subsection{Use-Case-Agnostic Offline Components} 
\label{sec:generic}

The use-case-agnostic components consume the load per system component (e.g., database, server, VM) and apply ML models to predict future load of this component. 

\textbf{Load Extraction Module} is implemented as a recurring query that extracts relevant data from raw production telemetry and stores this data in Azure Data Lake Store (ADLS)~\cite{data_lake}. These files are input to the AML pipeline.

For the backup scheduling scenario, we have selected the average customer CPU load percentage per five minutes as an indicator of customer activity. 
Other signals (memory, I/O, number of active connections, etc.) can be added to improve accuracy. In this paper, customer CPU load percentage per server is referred to as load per server for readability. Servers are due for full backup at least once a week. Thus, the load extraction query runs once a week per region. 



\textbf{AML Pipeline} is the core component of the \app\ infrastructure. It is built using the functionality of Azure Machine Learning (AML)~\cite{aml} that facilitates end-to-end machine learning life cycle. 
This pipeline consumes the load, validates it, extracts features, trains a model, deploys the model, and makes it accessible through a REST endpoint. The pipeline tracks the versions of deployed models, performs inference, and evaluates the accuracy of predictions. Results are stored in Cosmos DB~\cite{cosmosdb}, globally distributed and highly available database service. Based on predicted load, resource allocation can be optimized in various ways.

In our case, the predictions are input to the backup scheduling algorithm. A run of the AML pipeline is scheduled once a week per region since servers are due for full backup at least once a week. 
Due to space limitations, we describe five most interesting modules of the pipeline. They are:

$\bullet$ \textbf{\textit{Data Validation Module}}.
Since data validation is a well-studied topic~\cite{BPRWZ19}, we implemented existing rules such as detection of schema and bound anomalies. 

$\bullet$ \textbf{\textit{Feature Extraction Module}}. 
Lifespan and typical resource usage patterns are examples of the features that are useful for load prediction. In particular, we differentiate between short-lived and long-lived servers, stable and unstable servers, servers that follow a daily or a weekly pattern and servers that do not conform to such a pattern, predictable and unpredictable servers in Sections~\ref{sec:load} and \ref{sec:backup_scheduling_problem}.
We will extend this module by other features~\cite{PLT18} to improve accuracy.

$\bullet$ \textbf{\textit{Model Training and Inference Modules}}.
While many ML models can be plugged into the \app\ infrastructure, we compared NimbusML~\cite{nimbusml}, GluonTS~\cite{gluonts}, and Prophet~\cite{prophet} with respect to accuracy and scalability. We applied these models only to those servers that cannot be accurately predicted by persistent forecast in Section~\ref{sec:prediction}.

$\bullet$ \textbf{\textit{Accuracy Evaluation Module}}.
For the backup scheduling scenario, the accuracy of load prediction for the whole day per server is less critical than correct prediction of lowest load window per day per server. Thus, we tailor prediction accuracy to our use case. In particular, we measure if the lowest load window is chosen correctly and if the load during this window is predicted accurately in Sections~\ref{sec:load_prediction_accuracy} and \ref{sec:accuracy}. 


\textbf{Application Insights Dashboard}~\cite{appinsights} provides summarized view of the pipeline runs to facilitate real-time monitoring and incident management. Examples of incidents include missing or invalid input data, errors or exceptions in any step of the pipeline, and failed model deployment.

\subsection{Use-Case-Specific Online Components} 
\label{sec:specific}

The use-case-specific components leverage predicted load to optimize backup scheduling. The backup scheduler runs within Master Data Service (MDS) runner per day and cluster. The Runner Service deploys executables which probe their respective services resulting in measurement of availability and quality of service. The runner service is deployed in each Azure region.

For those servers that are due for full backups the next day, the backup scheduling algorithm verifies if these servers were predicted correctly for the last three weeks. This way, we verify that the servers were predictable for several weeks and we do not reschedule a backup at a worse time based on predictions we are not confident in. Three weeks of history is a compromise between prediction confidence and relevance of this rule to the majority of servers (58\% of servers survive beyond three weeks, Figure~\ref{fig:classificaiton}).
For such predictable servers, the algorithm extracts the predicted load for the next day and selects a time window during which customer activity is expected to be the lowest. The algorithm stores the start time of this window as a service fabric property of respective PostgreSQL and MySQL database instances. This property is used by the backup service to schedule backups. Servers that did not exist or were unpredictable for the last three weeks are scheduled for backup at default time. 

\subsection{Reuse of Seagull for Other Scenarios} 
\label{sec:reuse}

So far, we applied the \app\ infrastructure to two different scenarios: (1)~Backup scheduling of PostgreSQL and MySQL servers and (2)~Preemptive auto-scale of SQL databases (Appendix~\ref{sec:appendix}). Based on this experience, we now summarize how to reuse the use-case-agnostic components of \app.

\textbf{No Changes}.
All interfaces between the use-case-agnostic components, Model Deployment and Tracking are designed independently from any scenario and require no changes.

\textbf{Parameter Updates}.
Data Ingestion and Validation, storage of results to CosmosDB, Pipeline Scheduler, Incident Management, and Application Insights Dashboard are parameterized to facilitate easy adjustment to a new scenario. 
For example, to account for changes of input data, we automatically deduce schema and other data properties (e.g., min and max values of numeric attribute values) from the input data. The schema and data properties are stored in a file. After the file has been verified by a domain expert, it is used to detect schema and bound anomalies.

Other components require similar parameter updates. 
For example, Data Ingestion requires update of the location of input data in ADLS and access rights to this data.
Also, the schema of CosmosDB tables, frequency of pipeline runs, and gathered statistics may be different for other scenario.  

\textbf{Adjustments}.
Load Extraction, Feature Extraction, Model Training, Inference, and Accuracy Evaluation may require non-trivial customization. 
For example, other forecast signals (CPU, memory, disk, I/O, etc.) and features (subscriber identifier, number of active connections, etc.) may be needed for other scenarios. 
Accuracy and scalability of ML models heavily depends on the input data and scenario (Sections~\ref{sec:load} and \ref{sec:prediction}).
Accuracy Evaluation may have to be tailored to the use case requirements (Section~\ref{sec:accuracy}). 

%

%% file: sections/data.tex
\section{PostgreSQL and MySQL Servers}
\label{sec:load}


In this section, we first define load prediction accuracy metric and then use this metric to measure if a server has stable load or follows a daily or a weekly pattern. 

\subsection{Load Prediction Accuracy Metric}
\label{sec:load_prediction_accuracy}

While there are several established statistical measures of prediction error (e.g., mean absolute scaled error and mean normalized root mean squared error), we found them unintuitive and cumbersome to use in our case. They produce a number representing prediction error per server per day. They give no insights into whether the lowest load window was chosen correctly per server per day nor whether the load was predicted accurately during this window. Thus, Definitions~\ref{def:accurate_load_prediction} and \ref{def:correct_low_load} below define these two metrics.

\begin{figure}[!t]
\centering
\includegraphics[width=.7\columnwidth]{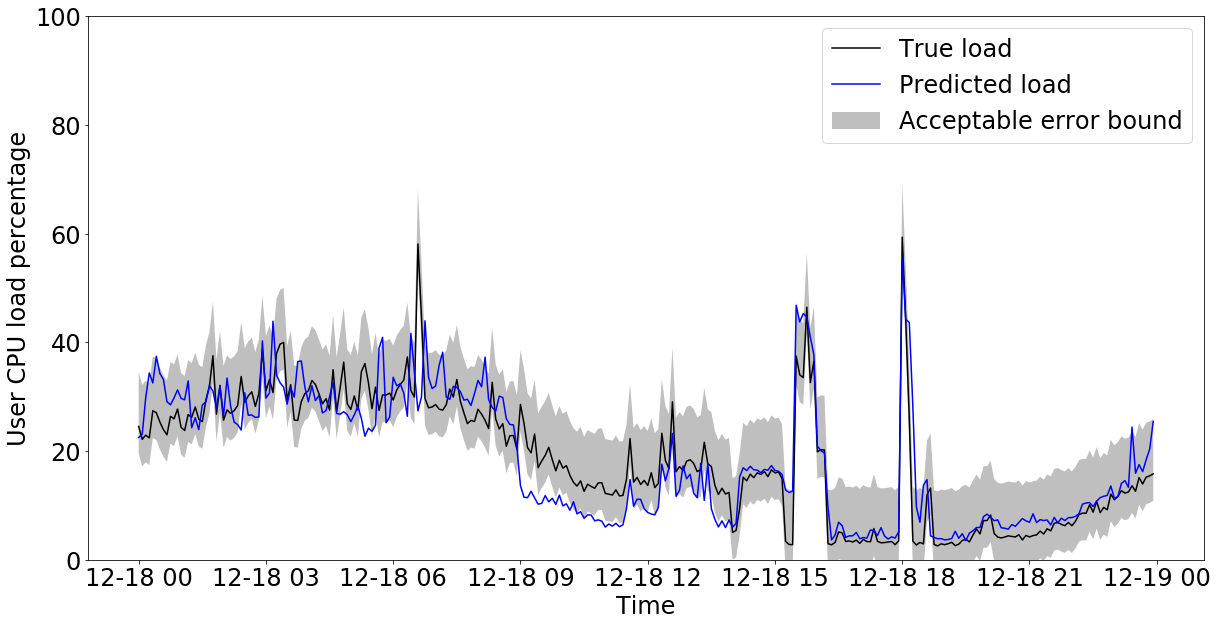}
\caption{Acceptable error bound}
\label{fig:error-bound}
\end{figure}

\begin{definition}(\textbf{Acceptable Error Bound, Bucket Ratio Metric})
Given predicted and true load for a server $s$ during a time interval $t$, we define the \textit{bucket ratio metric} of the server $s$ during the time interval $t$ as the percentage of predicted data points that are within the \textit{acceptable error bound} of +10/$-$5 of their respective true data points during the time interval $t$.
\label{def:bucket_ratio}
\end{definition}

Definition~\ref{def:bucket_ratio} specifies an asymmetric error bound that tolerates up to 10\% over-predicted load but only at most 5\% under-predicted load because a slight overestimation of low load periods is less critical for our use case than a slight underestimation that may result in interference with high customer load. 
In Definitions~\ref{def:bucket_ratio}--\ref{def:predictable}, we plug in constants that were empirically chosen by domain experts and are now used in production for the backup scheduling use case. Other constants can be plugged in for other scenarios. 


\begin{definition}(\textbf{Accurate Load Prediction})
Prediction of the load of a server $s$ during a time interval $t$ is \textit{accurate} if the bucket ratio of the server $s$ during the time interval $t$ is at least 90\%. Otherwise, a prediction is \textit{inaccurate}.
\label{def:accurate_load_prediction}
\end{definition}

In Figure~\ref{fig:error-bound}, we depict predicted load as blue line, true load as back line, and acceptable error bound as gray shaded area. Intuitively speaking, a prediction is accurate if 90\% of the blue line is in the shaded area. Even though for a human eye the prediction looks ``close enough", the bucket ratio is only 75\% and thus this prediction is inaccurate. This example illustrates that Definitions~\ref{def:bucket_ratio} and \ref{def:accurate_load_prediction} impose quite strict constraints on prediction accuracy.

\subsection{Server Classification}
\label{sec:classification}

We classify the servers with respect to their lifetime and typical customer activity patterns in Figure~\ref{fig:classificaiton}.
The classification provides us valuable insights about load predictability per class of servers. We will leverage these insights while choosing the ML model in Section~\ref{sec:prediction}. 

Given a random sample of several tens of thousands of servers from four regions during one month in 2019, Figure~\ref{fig:classificaiton} summarizes the percentage of servers that belong to each class. We define each class of servers below.

\subsubsection{Server Lifespan}

Servers are classified into short-lived and long-lived.

\begin{definition}(\textbf{Short-Lived Server})
%
A server is called \textit{long-lived} if it existed more than three weeks. 
Otherwise, a server is called \textit{short-lived}.
\label{def:lifetime}
\end{definition}

\begin{figure}[!t]
\centering
\includegraphics[width=.6\columnwidth]{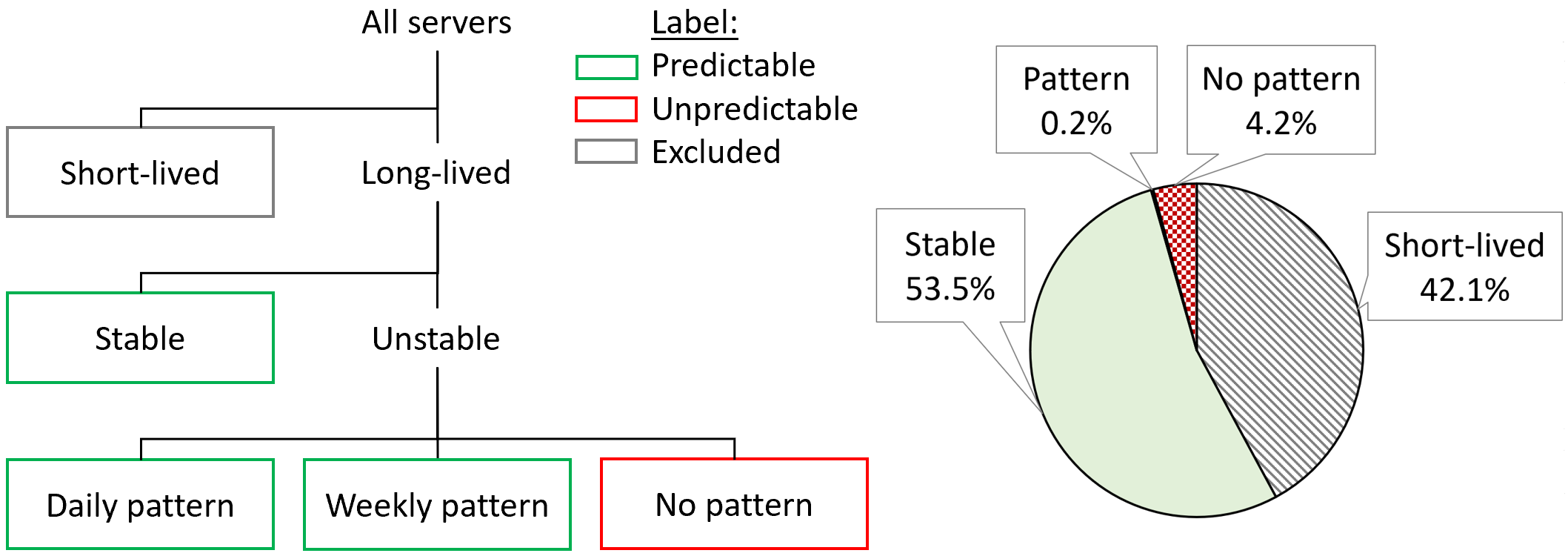}
\caption{Classification of servers}
\label{fig:classificaiton}
\end{figure}

As shown in Figure~\ref{fig:classificaiton}, 58\% of servers ``survive" for more than three weeks creating enough history to make a reliable conclusion whether they are predictable or not (Section~\ref{sec:backup_scheduling_problem}).
Remaining 42\% of servers are short-lived (Figure~\ref{fig:classificaiton}). We exclude them from further consideration.

\subsubsection{Typical Customer Activity Patterns}

We differentiate between stable and unstable servers. 

\begin{definition}(\textbf{Stable Server})
A long-lived server is called \textit{stable} during a time interval $t$ if its load is accurately predicted by its average load during the time interval $t$ (Definition~\ref{def:accurate_load_prediction}). 
Otherwise, a server is called \textit{unstable}.
\label{def:stable}
\end{definition}

\begin{figure}[!t]
\centering
\includegraphics[width=.7\columnwidth]{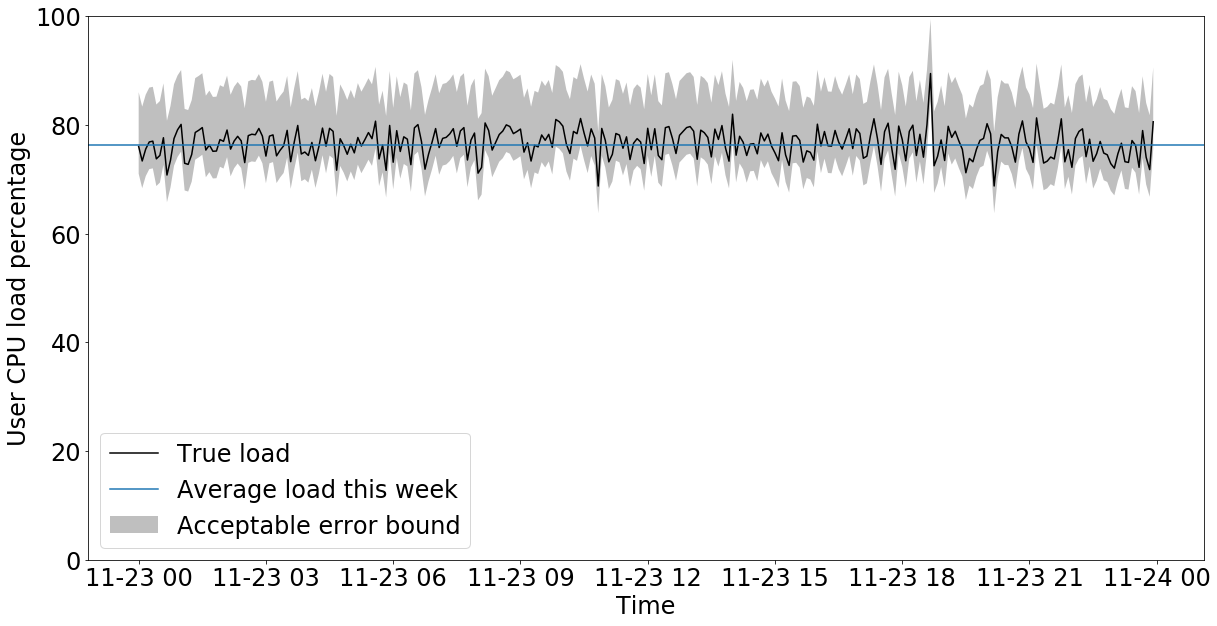}
\caption{Stable server}
\label{fig:stable}
\end{figure}

Figure~\ref{fig:stable} shows the true load of a server as a black line, the average load of this server during this week as a blue line, and the acceptable error bound as shaded gray area. The load of this server during this week is stable since the blue line is almost completely within the gray area. The bucket ratio is 99\% for this server on this week (Definition~\ref{def:bucket_ratio}). 

53.5\% of servers are long-lived and stable and thus easily predictable (Figure~\ref{fig:classificaiton}). 4.4\% of long-lived unstable servers require a more detailed analysis. They are further classified into those that follow a daily or a weekly pattern and those that do not conform to such a pattern.

\begin{definition}(\textbf{Server with Daily Pattern})
Given the load of a server $s$ on two consecutive days $d-1$ and $d$, the server $s$ has a \textit{daily pattern} on day $d$ if its load on day $d$ is accurately predicted by its load on the previous day $d-1$. 

A server has a \textit{daily pattern} during a time interval $t$ if its load conforms to this daily pattern on each day during the whole time period $t$. 
\label{def:daily-pattern}
\end{definition}

\begin{figure}[!t]
\centering
\includegraphics[width=.7\columnwidth]{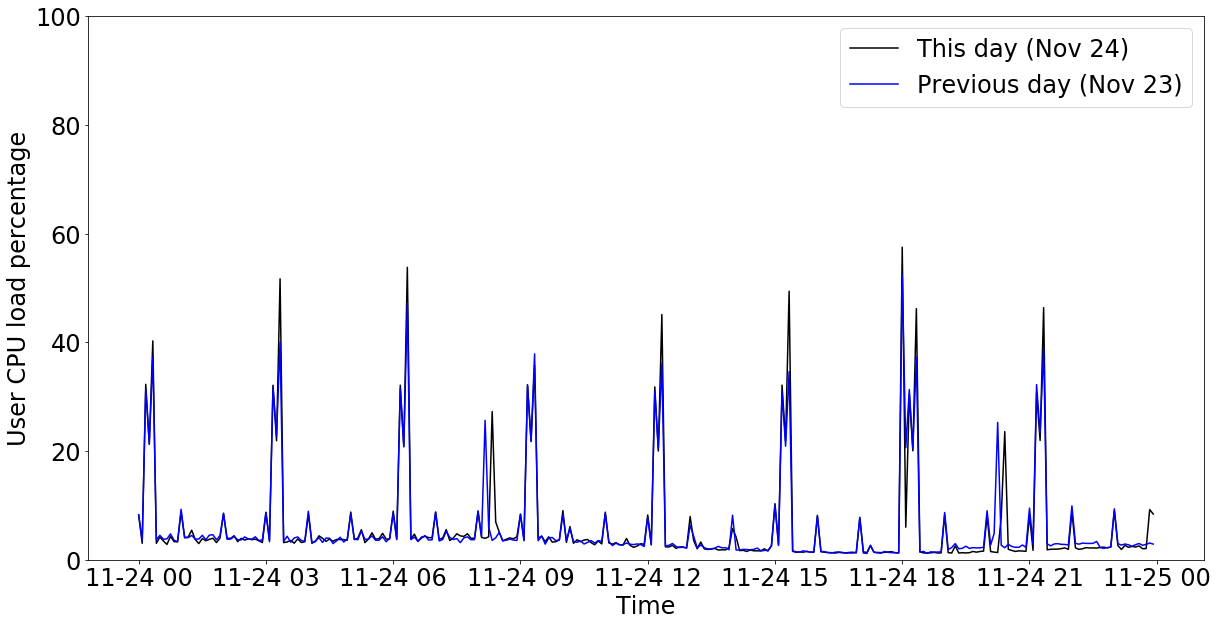}
\caption{Server with daily pattern}
\label{fig:daily_pattern}
\end{figure}

Figure~\ref{fig:daily_pattern} shows an example of a server with a strong daily pattern. We plot the load on this day in black and on the previous day in blue. These lines overlap almost perfectly. The bucket ratio is 95\%. Such a precise daily pattern could be the result of an automated recurring workload.

\begin{definition}(\textbf{Server with Weekly Pattern})
Given the load of a server $s$ on two consecutive equivalent days of the week $d-7$ and $d$, the server $s$ has a \textit{weekly pattern} on day $d$ if its load on day $d$ is accurately predicted by its load on the previous equivalent day of the week $d-7$. 

A server has a \textit{weekly pattern} during a time interval $t$ if it does not have a daily pattern during the time period $t$ and its load conforms to a weekly pattern on each day during the whole time interval $t$.
\label{def:weekly-pattern}
\end{definition}

\begin{figure}[!t]
\centering
\includegraphics[width=.7\columnwidth]{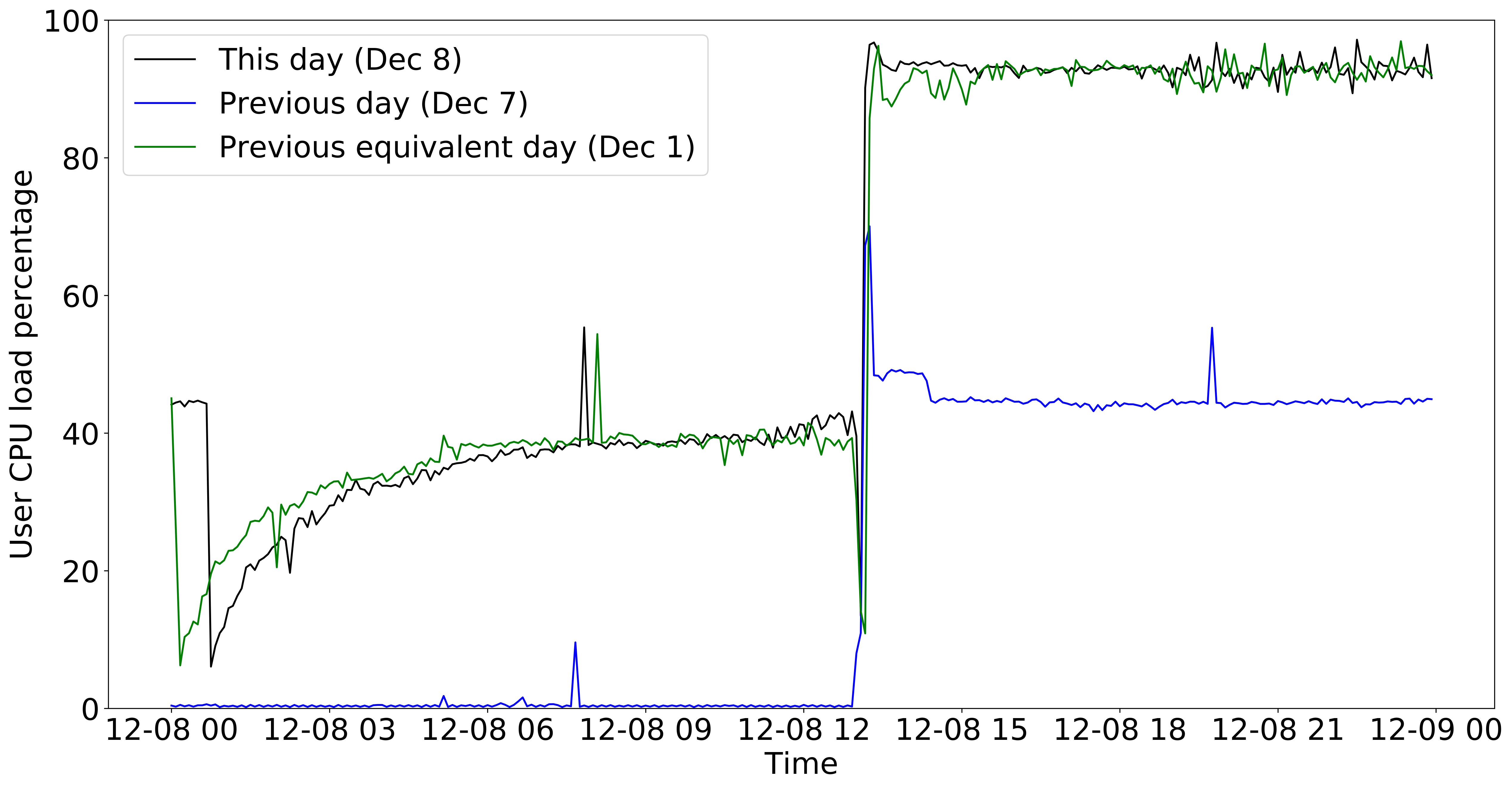}
\caption{Server with weekly pattern}
\label{fig:weekly_pattern}
\end{figure}

Figure~\ref{fig:weekly_pattern} shows an example of a server that follows a weekly pattern. Similarly to previous Sunday (December 1), the load on this Sunday (December 8) is medium before noon and high after noon. The bucket ratio is over 90\%.
In contrast, the load on previous day (December 7) is low before noon and medium after noon. The bucket ratio is only 1\%.
Thus, we conclude that this server follows a weekly pattern but does not conform to a daily pattern.

0.2\% of servers conform to a daily or a weekly pattern and thus are easy to predict (Figure~\ref{fig:classificaiton}). Even though this percentage is relatively low, hundreds of top-revenue customers fall into this class of servers and cannot be disregarded.

\begin{figure}[!t]
\centering
\includegraphics[width=.7\columnwidth]{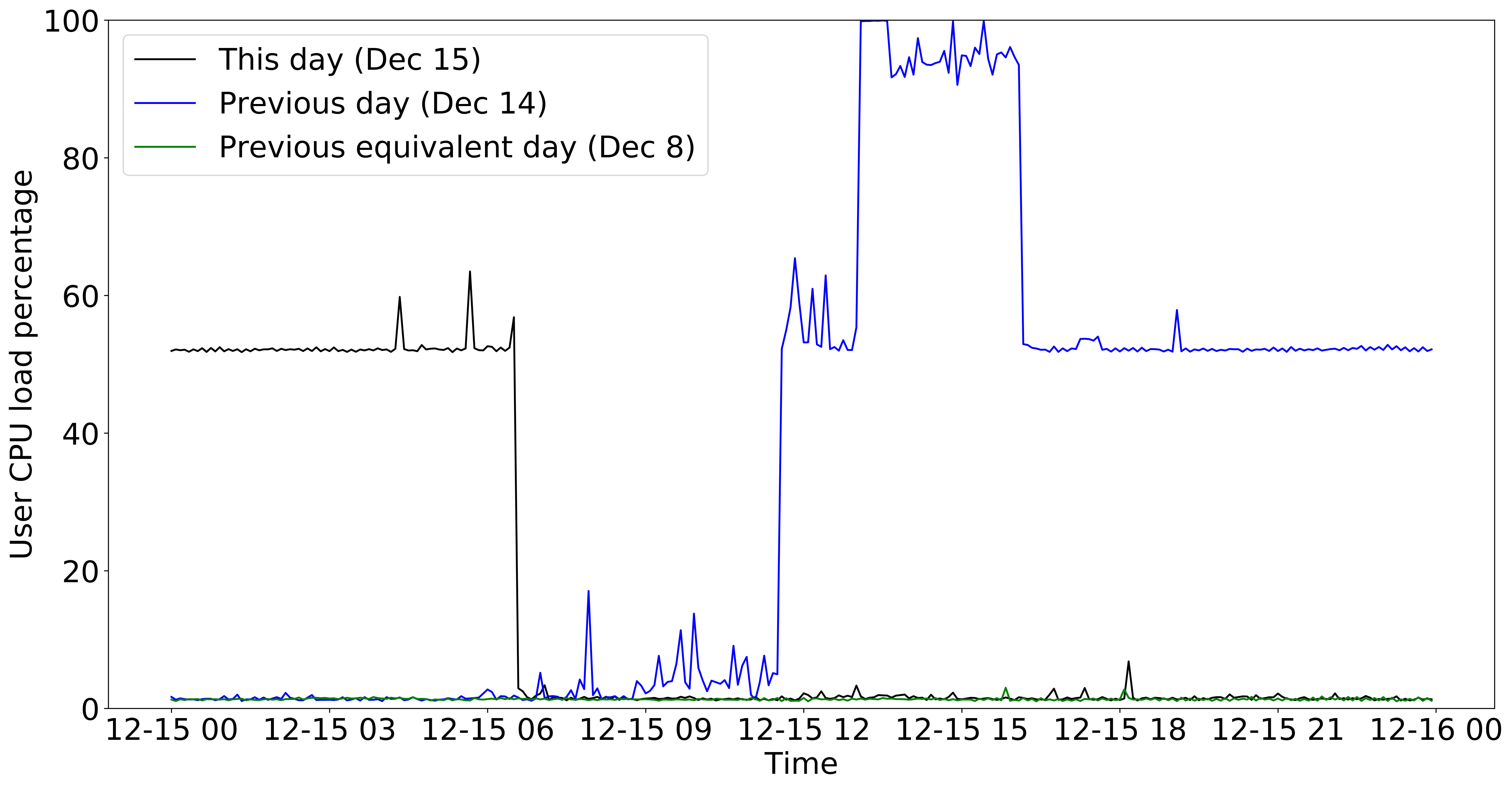}
\caption{Server without daily or weekly pattern}
\label{fig:without_pattern}
\end{figure}

Figure~\ref{fig:without_pattern} illustrates the load of a server without any daily or weekly pattern. User idle time after 6AM was expected since the user was idle at the same time on the previous equivalent day (i.e., previous Sunday). However, high user activity before 6AM was not typical for this server neither the day before nor on the previous Sunday.
The bucket ratio based on the previous day is 20\% and based on the previous equivalent day is 72\%. That is, this server follows neither daily, nor weekly pattern.
4.2\% of servers do not have any pattern. They tend to be unpredictable (Section~\ref{sec:backup_scheduling_problem}). 

\textbf{Summary}.
Figure~\ref{fig:classificaiton} illustrates that 53.7\% of servers is expected to be predictable because their load is either stable or conforms to a pattern. 
4.2\% of the servers are neither stable nor follow a pattern. They are likely to be unpredictable. 
42.1\% are short-lived and thus excluded from further consideration. 
These insights will be used while choosing the ML model to predict low load per server in Section~\ref{sec:prediction}.

%% file: sections/accuracy.tex
\section{Low Load Prediction Accuracy}
\label{sec:accuracy}

In addition to the load prediction accuracy metric in Section~\ref{sec:load_prediction_accuracy}, we now define the lowest load window metric. Based on these metrics, we then formulate the backup scheduling problem statement.

\subsection{Lowest Load Window Metric}

For each server on its backup day, our goal is to predict the lowest valley in the user load that is long enough to fit a full backup of this server. The time interval of this valley is called the lowest load window. We  measure if this window is chosen correctly and if the load during this window is predicted accurately. Accurate prediction of the load during the rest of the day is less critical for our purposes.

\begin{definition}(\textbf{Lowest Load (LL) Window})
Let $s$ be a server which is due for full backup on day $d$.
Let $b$ be the expected duration of full backup of the server $s$.
\textit{True LL window} for the server $s$ on the day $d$ is the time interval of length $b$ during which the average true load of the server $s$ on the day $d$ is minimal across all other time intervals of length $b$ on the day $d$. 
\textit{Predicted LL window} is defined analogously based on predicted load of the server $s$ on day $d$.
\label{def:ll_window}
\end{definition}

\begin{definition}(\textbf{Correctly Chosen LL Window})
Let $w_t$ and $w_p$ be the true and predicted LL windows for a server $s$ on day $d$.
If the average true load during the predicted LL window $w_p$ is within an acceptable error bound of the average true load during the true LL window $w_t$, we say that the predicted LL window $w_p$ is \textit{chosen correctly}.
\label{def:correct_low_load}
\end{definition}

\begin{figure}[!t]
\centering
\includegraphics[width=.7\columnwidth]{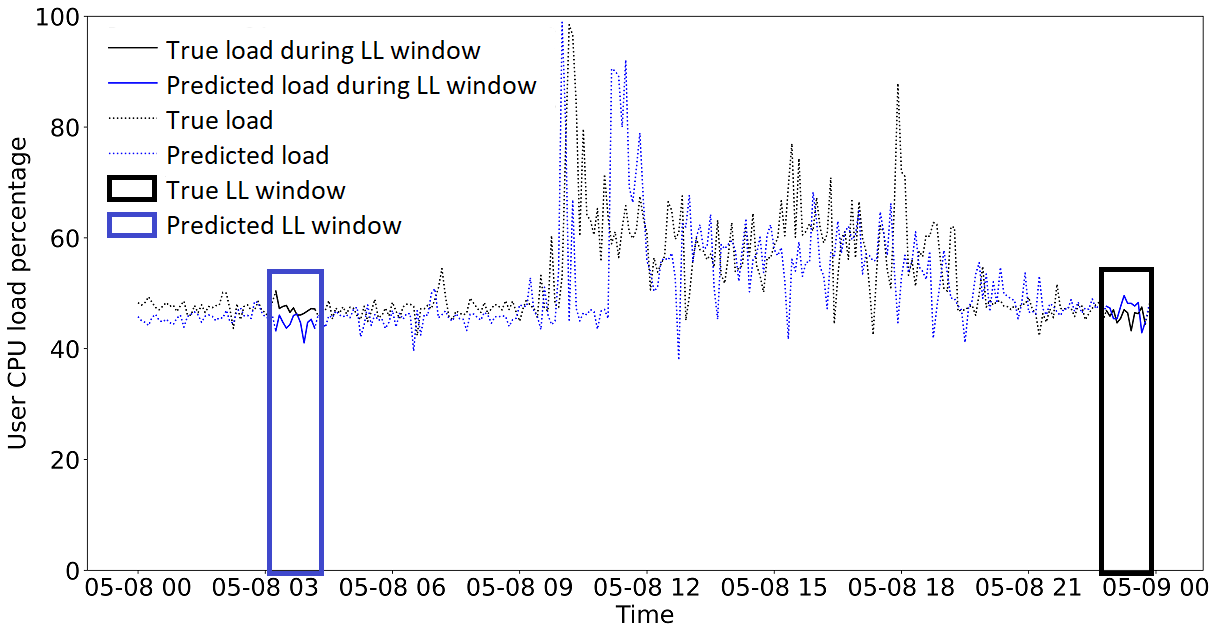}
\caption{Correctly chosen LL window}
\label{fig:correctly_chosen_window}
\end{figure}

\begin{figure}[!t]
\centering
\includegraphics[width=.7\columnwidth]{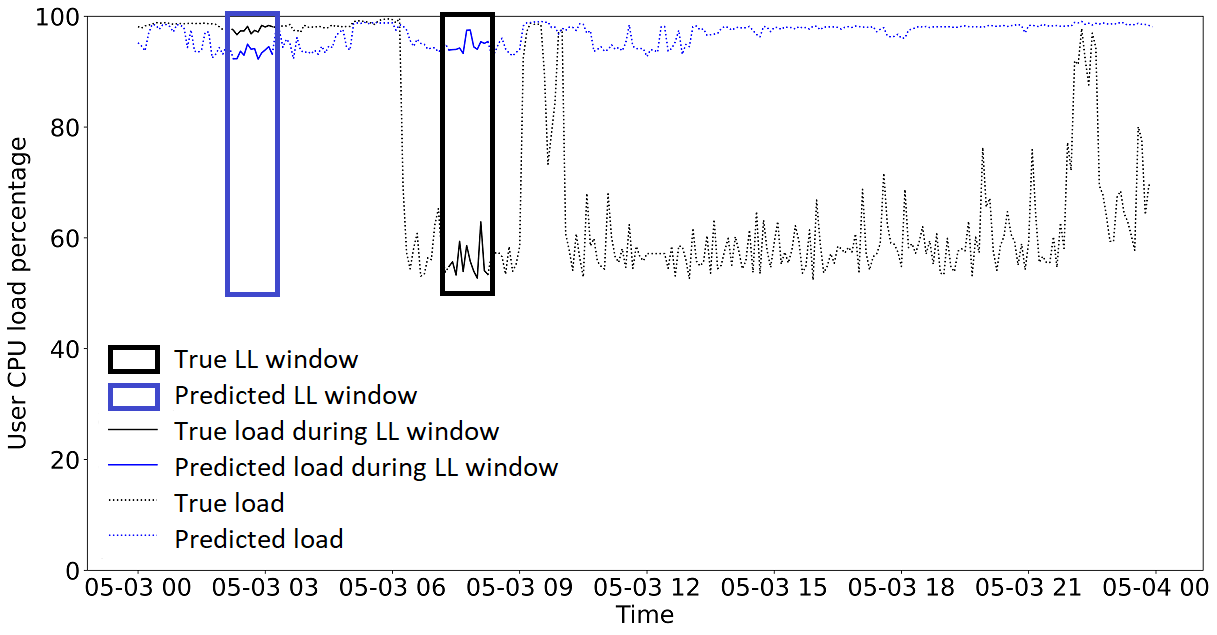}
\caption{Incorrectly chosen LL window}
\label{fig:wrong_window}
\end{figure}

In Figure~\ref{fig:correctly_chosen_window}, the true and predicted LL windows do not overlap.
%
%
However, the average true load during true LL window is only slightly lower than the average true load during predicted LL window. Therefore, the true LL window would not be a significantly better time interval to run a backup than the predicted LL window. Hence, we conclude that the predicted LL window is chosen correctly.


\subsection{Backup Scheduling Problem Statement}
\label{sec:backup_scheduling_problem}

In Section~\ref{sec:prediction}, we focus on one instance of the load prediction problem (Section~\ref{sec:infrastructure}).
Namely, for each server $s$ that is due for full backup on day $d$, our \app\ approach aims to:
(1)~Correctly choose the LL window on day $d$ to schedule a backup during this LL window on day $d$ and 
(2)~Accurately predict the load during this LL window to move a backup from default backup day to another day of the week if the load is lower on another day.

\begin{figure}[!t]
\centering
\includegraphics[width=.7\columnwidth]{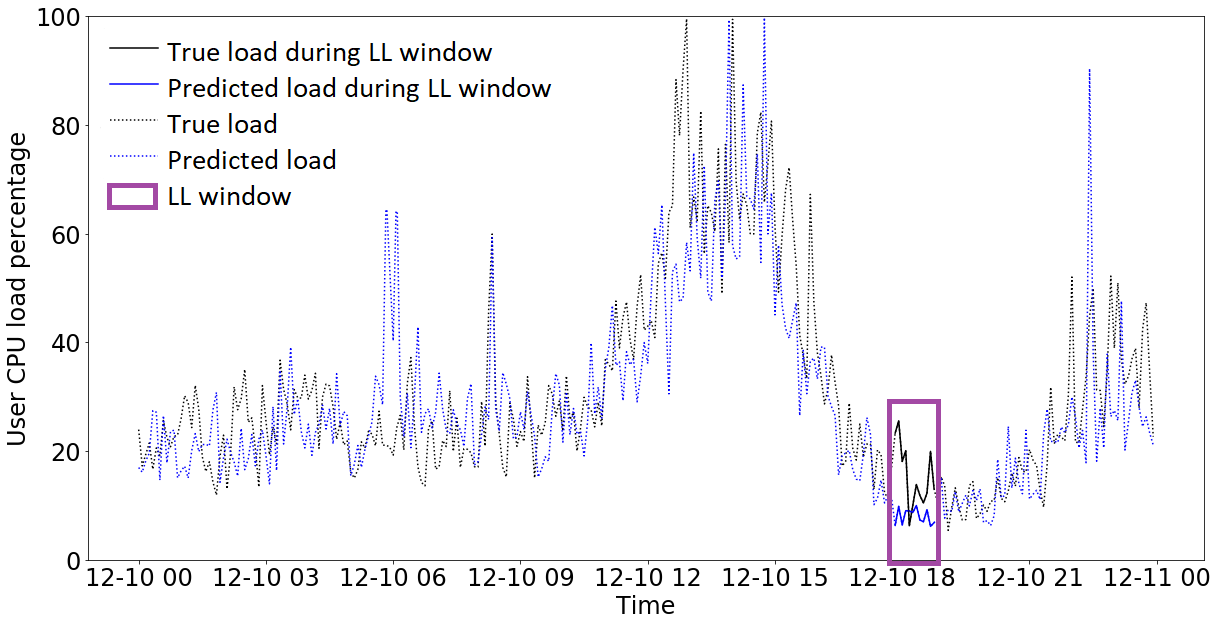}
\caption{Low load prediction accuracy}
\label{fig:inaccurate_load}
\end{figure}

These two metrics are orthogonal. 
For example, the true and predicted LL windows coincide in Figure~\ref{fig:inaccurate_load}. Thus, the LL window is chosen correctly. However, the load prediction during this window is not accurate. Indeed, the true load is significantly higher than the predicted load and the bucket ratio is only 50\% during this window. 

The opposite case is also possible. Namely, the true load is predicted accurately during predicted LL window in Figure~\ref{fig:wrong_window}. The bucket ratio is 92\%. However, the true load during the true LL window is much lower than during the predicted LL window. Thus, the LL window is not chosen correctly in this case.
%
Based on these observations, we conclude that only both metrics combined give us reliable insights about low load prediction accuracy. 


\begin{definition}(\textbf{Predictable Server})
A long-lived server is called \textit{predictable} if for the last three weeks its LL windows were chosen correctly and the load during these windows was predicted accurately (Definitions~\ref{def:accurate_load_prediction} and \ref{def:correct_low_load}).
%
\label{def:predictable}
\end{definition}

As explained in Section~\ref{sec:infrastructure}, we change backup window for predictable servers only. Servers that did not exist or were not predictable for three weeks, default to current backup time that is chosen independently from customer activity.

%% file: sections/ml.tex
\section{Low Load Prediction}
\label{sec:prediction}

\input{sections/accuracy_evaluation}

In this section, we first describe the ML models that are commonly used for time series forecast and then choose a model per each class of servers and compare the models with respect to their accuracy and scalability.

\subsection{ML Models for Time Series Forecast}
\label{sec:models}

We now summarize the key ideas of the ML models that we considered to predict the low customer activity per server on its backup day.These models range from simple heuristics to complex neural-network-based ones. 

\textbf{Persistent Forecast} refers to replicating previously seen load per server as the forecast of the load for this server. We compared three variations of the persistent forecast model:

$\bullet$ \textit{Previous day} takes the load a server on the previous day and utilizes it as predicted load on the next day.

$\bullet$ \textit{Previous equivalent day} forecasts the load of a server by replicating its load on previous equivalent day of the week. 

$\bullet$ \textit{Previous week average} makes a prediction as the average load of a particular server during previous week. 

\textbf{NimbusML}~\cite{nimbusml} is a Python module that provides Python bindings for ML.NET. NimbusML aims to enable data science teams that are more familiar with Python to take advantage of ML.NET's functionality and performance. It provides battle-tested, state-of-the-art ML algorithms, transforms, and components. Specifically, we use Singular Spectrum Analysis to transform forecasts.

\textbf{GluonTS}~\cite{gluonts} is a toolkit for probabilistic time series modeling, focusing on deep learning-based models. We train a simple feed forward estimator. We tried several other estimators but this model achieved highest accuracy. 

\textbf{Prophet}~\cite{prophet} is open source software released by Facebook. It forecasts a time series data based on an additive model where non-linear trends are fit with yearly, weekly, and daily seasonality, plus holiday effects. It works well for time series that have strong seasonal effects and several seasons of historical data. Prophet is robust to missing data and shifts in the trend, and typically handles outliers well. 

\textbf{ARIMA}: Auto-Regressive Integrated Moving Average mo\-del~\cite{arima} forecasts the future values of a series based on the different seasonal and temporal structures in the series. At inference, it predicts one signal at a time by fitting to this signal's prior values. 

\subsection{ML Model per Class of Servers}
\label{sec:choice-of-model}

In this section, we discuss the applicability of each model to each class of servers we identified in Section~\ref{sec:classification}. We differentiate between two cases:

$\bullet$ \textit{Stable servers and servers that follow business patterns that can be recognized by persistent forecast}. Obviously, such servers can be accurately predicted by persistent forecast and no complex ML models are needed.
Indeed, the previous week average can predict the load of stable servers (Definition~\ref{def:stable}); 53.5\% of servers are stable (Figure~\ref{fig:classificaiton}). 
%
Previous equivalent day is more powerful than previous week average because it captures a weekly pattern (Definition~\ref{def:weekly-pattern}), including stable load
which covers 53.6\% of servers. 
%
Previous day is also more powerful than previous week average, since it captures a daily pattern (Definition~\ref{def:daily-pattern}), including stable loads. 
53.7\% of servers can be predicted by the previous day's pattern. 
%
%
Since previous day is suitable for the largest subset of servers, we focus on this variant in the following.

$\bullet$ \textit{Unstable servers that do not conform to a pattern that can be recognized by persistent forecast}. 
4.2\% of servers fall into this category. 
In Section~\ref{sec:comparison}, we apply ML models to such servers to find out if these models can detect a predictable load pattern for these servers.

\subsection{Experimental Comparison of ML Models}
\label{sec:comparison}

\subsubsection{Experimental Setup}
\label{sec:setup}

\textbf{Hardware}.
We conducted all experiments on a VM running Ubuntu 18.04. It has 16 CPUs and 64GB of memory.

\textbf{Input Data}.
As described in Section~\ref{sec:infrastructure}, the pipeline runs per Azure region once a week. Thus, our input data is partitioned by region and week. Since the size of regions varies, the size of input files ranges from hundreds of kilobytes to a few gigabytes. Below, we randomly selected four input files with different sizes to demonstrate the scalability of ML models and find out if there are differences in accuracy of predictions between models and regions. The input files are in csv format. They contain server identifier, timestamp in minutes, average user CPU load percentage per five minutes, default backup start and end timestamps. 

In order to identify predictable servers, we have to consider three weeks (Definition~\ref{def:predictable}). To infer the load per server on its backup day, ML models are trained on one week of data prior to backup day per server. Thus, each input data set contains four weeks in one region, unless stated otherwise. 
We consider servers have at least three days of history prior to their backup days to train the ML models. 

\textbf{Methodology}.
We implemented the \app\ pipeline in Python. 
Our base-line implementation is \textit{single-thread\-ed}. 
%
%
Our \textit{multi-threaded} Dask-based~\cite{dask} implementation partitions the data per server and processes servers in parallel. 

\textbf{Metrics}.
For each ML model, we measure the percentage of correctly chosen LL windows, the percentage of LL windows with accurately predicted load, and the percentage of predictable servers among servers that existed at least three weeks (Definitions~\ref{def:accurate_load_prediction}, \ref{def:correct_low_load}, and \ref{def:predictable}). 
We measure the runtime of training, inference, and accuracy evaluation in minutes. 

\subsubsection{Stable Servers and Servers with Pattern}
\label{sec:stable}

As explained in Section~\ref{sec:choice-of-model}, majority of long-lived servers have stable load or follow daily or weekly patterns that can be recognized by persistent forecast. Therefore, we use persistent forecast to predict the load of such servers. 
For our sample data set, this heuristic correctly selected 99.83\% of LL windows, accurately predicted the load during 99.06\% of all windows, and classified 96.92\% of servers as predictable.

\subsubsection{Unstable Servers Without Pattern}
\label{sec:unstable}

We now apply ML models from the tools mentioned in Section~\ref{sec:models} to unstable servers that do not follow business patterns that can be recognized by persistent forecast. 

\textbf{Training and Inference}.
\textit{Persistent forecast} does not require training because it uses the load per server on the previous day as predicted load per server on the next day.

\textit{NimbusML} scales well (Figure~\ref{fig:training_inference_unstable}). Runtime for training and inference increases linearly from 2.5 seconds to 4 minutes as the number of servers grows from 10 to 700. Some of these measurements are not visible due to log scale with base 10 in Figure~\ref{fig:training_inference_unstable}.

\textit{GluonTS} also scales well.  
Training time ranges from 4 to 10 minutes, while inference time ranges from 0.2 to 16 seconds as the number of servers grows from 10 to 700.

\textit{Prophet} does not scale as well. Its training time grows from 1 to 34 minutes, while inference takes from 1 to 15 hours as the number of servers increases from 10 to 100. 
Thus, we implemented Prophet on Dask and achieved up to 60X speedup compared to single-threaded execution. 
However, when the number of servers exceeds 200, Prophet runs out of memory on Dask independently from the number of workers. Single-threaded execution does not terminate. 

\textit{ARIMA} is computationally intensive since it searches the optimal values of six parameters per server in order to make an accurate load prediction per server. We had explored parameter sharing between servers but that resulted in a worsening of accuracy. While inference time is within a few seconds per server, fitting may take up to 3 hours per server. Hence, executing ARIMA in parallel for each server does not make runtime of ARIMA comparable to other models. 

\textbf{Low Load Prediction Accuracy}.
%
%
NimbusML correctly chooses the highest percentage of LL windows compared to other tools (Figure~\ref{fig:ll_window_unstable}).
There is slight variance in accuracy of load prediction during LL windows and the percentage of predictable servers across regions and models (Figures~\ref{fig:load_unstable} and \ref{fig:predictable_unstable}).
Accuracy of persistent forecast, NimbusML, and GluonTS is comparable with respect to these two metrics. 
Prophet has similar or lower accuracy compared to the other two tools.


\subsection{Choice of Model for Final Deployment}

To find the middle ground between the accuracy of low load prediction and the overhead of model training and inference, we deployed persistent forecast based on previous day to production.
This heuristic correctly selected 99\% of low load windows, accurately predicted the load during 96\% of all windows, and classified 75\% of long-lived servers as predictable. The accuracy of other models is not significantly higher than the accuracy of persistent forecast.
Persistent forecast does not introduce any computational delay due to training and thus scales better than other models.
Lastly, it is easier to maintain a single model for the entire fleet of servers than a different model per each class of servers.

%% file: sections/accuracy_evaluation.tex

\begin{figure*}[!t]
\centering
    \subfigure[Training and inference]{
    	\includegraphics[width=0.3\columnwidth, keepaspectratio]{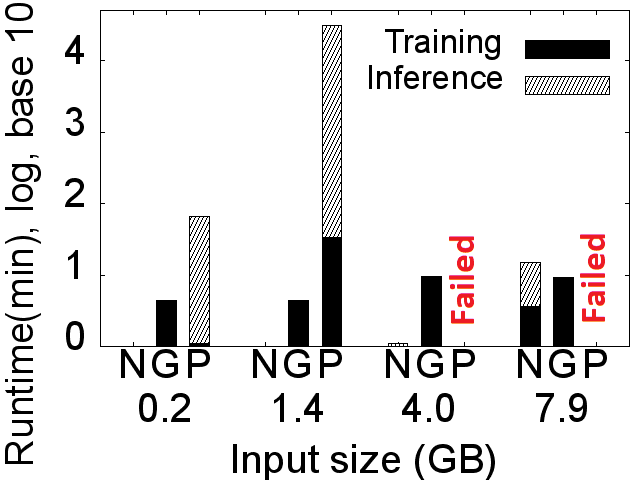}
    	\label{fig:training_inference_unstable}
	}\hspace{2cm}
	\subfigure[LL windows]{
    	\includegraphics[width=0.3\columnwidth, keepaspectratio]{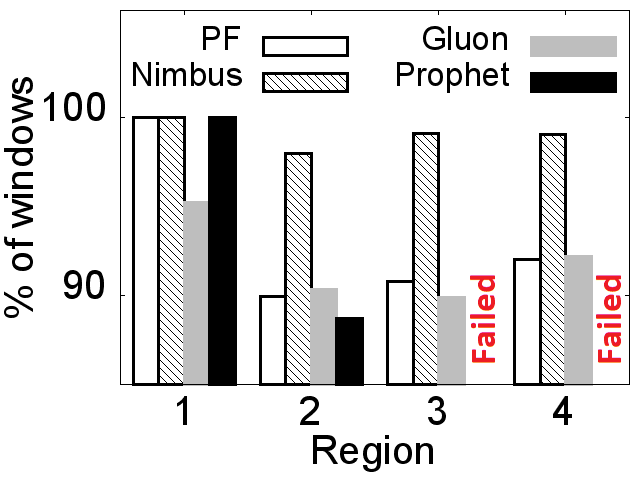}
    	\label{fig:ll_window_unstable}
	}\\
	\subfigure[Load during LL windows]{
    	\includegraphics[width=0.3\columnwidth, keepaspectratio]{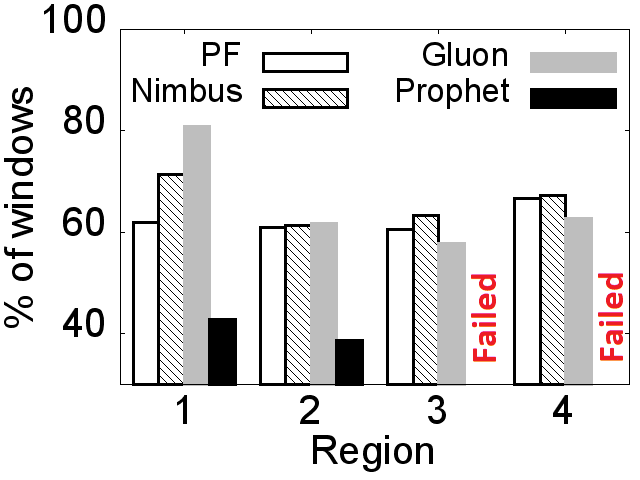}
    	\label{fig:load_unstable}
	}\hspace{2cm}
	\subfigure[Predictable servers]{
    	\includegraphics[width=0.3\columnwidth, keepaspectratio]{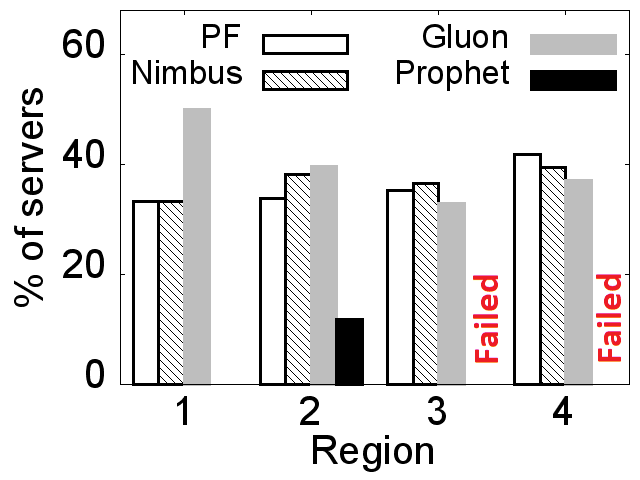}
    	\label{fig:predictable_unstable}
	}
\caption{Low load prediction using Persistent Forecast (PF), Nimbus (N), Gluon (G), and Prophet (P)}
\label{fig:exp_unstable}
\end{figure*}

%% file: sections/evaluation.tex
\section{Infrastructure Evaluation}
\label{sec:evaluation}

In this section, we evaluate runtime, scalability, and impact of the \app\ infrastructure.

\subsection{Runtime and Scalability}

\input{sections/runtime-evaluation}

In Figure~\ref{fig:runtime-all}, we measure the runtime of the use-case-agnostic components per Azure region. These components are: Data Ingestion, Data Validation, Feature Extraction, Model Deployment, and Accuracy Evaluation.
Runtime of Model Training and Inference per ML model are evaluated in Figure~\ref{fig:training_inference_unstable}.
Model Tracking, Pipeline Scheduler, and Incident Management run concurrently with other components and do not block the flow of the data through the AML pipeline. Thus, they are omitted in Figure~\ref{fig:runtime-all}. 
Since Load Extraction runs outside of the pipeline for all regions at once, it is also omitted. Load Extraction takes 30 minutes given the petabyte scale and complexity of raw telemetry.

In Figure~\ref{fig:exp_runtime}, we measure the runtime for the same four regions of different sizes as in Figure~\ref{fig:training_inference_unstable}.
While Figure~\ref{fig:training_inference_unstable} considers four weeks to train ML models and infer future load, Figure~\ref{fig:exp_runtime} considers only one week which corresponds to our production settings (Section~\ref{sec:generic}). 

Model Deployment takes about one minute independently from deployed model and input data size. In contrast, runtime of other components increases linearly with growing input size. 
When input size exceeds 1GB, Accuracy Evaluation becomes a bottleneck. Thus, we partitioned input data per server and ran Accuracy Evaluation in parallel per server using Dask~\cite{dask}. Figure~\ref{fig:runtime-dask} compares single-threaded and multi-threaded Accuracy Evaluation per server on its backup day. While Dask is 5 seconds slower than the single-threaded execution for 60MB, Dask consistently wins for input sizes over 400MB. For 2.5GB, Dask is 26\% faster than single-threaded execution.

To further optimize backup scheduling, we will move a backup of a server from its default backup day to other day of the week if the load is lower and/or prediction is more accurate on another day. 
In Figure~\ref{fig:runtime-dask}, we also measure the runtime of accuracy evaluation on each day one week ahead per server. Dask consistently achieves 3-4.6X speedup compared to the single-threaded implementation for all input sizes. For 2.5GB, the single-threaded implementation runs for over 1 hour, while Dask terminates after 15 minutes which we consider to be an acceptable computational delay for a large Azure region.

\subsection{Impact and Future Work}
\label{sec:impact}

\input{sections/impact-evaluation}

The impact of \app\ is two-fold, namely, it improves customer experience and reduces engineering effort.

\textbf{Improving Customer Experience}.
The \app\ infrastructure is deployed for tens of thousands of PostgreSQL and MySQL servers in tens of Azure regions to optimize backup scheduling.
In Figure~\ref{fig:backups}, we compare predicted LL windows  (Definition~\ref{def:ll_window}) to default backup windows for all servers in all regions during one month in 2020. 

For busy servers with customer load over 60\% of capacity, 7.7\% of backup collisions with peaks of customer activity are now avoided which corresponds to several hundred hours of improved customer experience. 
91.1\% of default windows are as good as LL windows on respective backup days. This happens by chance when default windows do not collide with high customer load.
Only 1.2\% of LL windows were not chosen correctly. This can be explained by unexpected change of customer behavior compared to the previous three weeks (Definition~\ref{def:predictable}).
For servers with predictable daily patterns (Definition~\ref{def:daily-pattern}), the results are similar. In particular, 12.5\% of backups were moved from default windows that coincided with customer activity to correctly chosen LL windows (Definition~\ref{def:correct_low_load}). This percentage corresponds to several hundred hours of improved customer experience. 
As expected, for stable servers (Definition~\ref{def:stable}), 99.5\% of default windows correspond to LL windows. 

We also use the lowest load window metric (Section~\ref{sec:accuracy}) to measure if backup windows selected by customers correspond to predictable lowest load windows and suggest windows with expected lower load instead. 

While analysing the load, we concluded that many servers are not only predictable but also do not use the full capacity most of the time. 
Figure~\ref{fig:autoscale} illustrates the percentage of servers per maximal CPU load percentage of capacity per time unit. 
Only 3.7\% of servers reach their CPU capacity per week, i.e., for 96.3\% of servers resources could be saved.
This observation opens up opportunities to overbook or auto-scale resources~\cite{LRDXGKC16, RDG11}. We will explore these optimization techniques in follow-up projects to amplify impact.

\textbf{Reducing Engineering Effort}.
Thanks to the automated workload analysis enabled by \app, the engineers do not have to manually study customer behavior to select backup windows (Section~\ref{sec:introduction}). This approach was labor-intensive, time-consuming, neither scalable to millions of customers, nor durable since load patterns change over time.

Based on the \app\ infrastructure, the time to setup a load prediction pipeline for other use cases came down from months to weeks. As described in Section~\ref{sec:reuse}, we applied \app\ to two scenarios so far. It took several months for a dedicated team of three software engineers, two data scientists, and a project manager to build, optimize, test, and deploy the \app\ infrastructure to production worldwide. However, it took only a few weeks to adjust this infrastructure to a new use case. In particular, we updated parameters of the use-case-agnostic components, re-implemented Load Extraction and Accuracy Evaluation, and hooked the predictions with the backup scheduling service.


%% file: sections/runtime-evaluation.tex
\begin{figure}[!t]
\centering
	\subfigure[All components]{
    	\includegraphics[width=0.3\columnwidth, keepaspectratio]{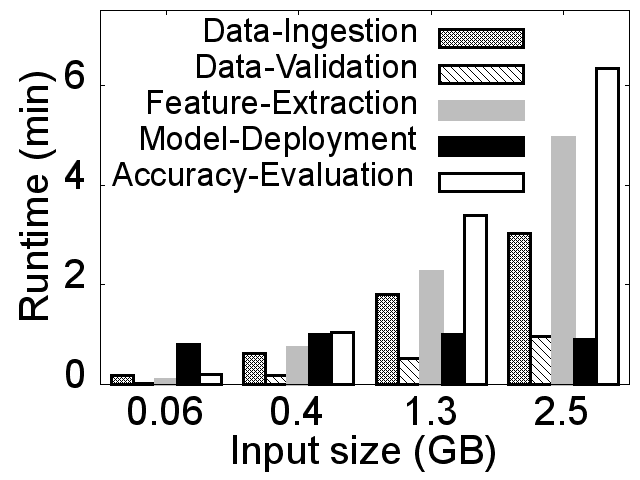}
    	\label{fig:runtime-all}
	}\hspace{2mm}
	\subfigure[Accuracy~Evaluation]{
    	\includegraphics[width=0.3\columnwidth, keepaspectratio]{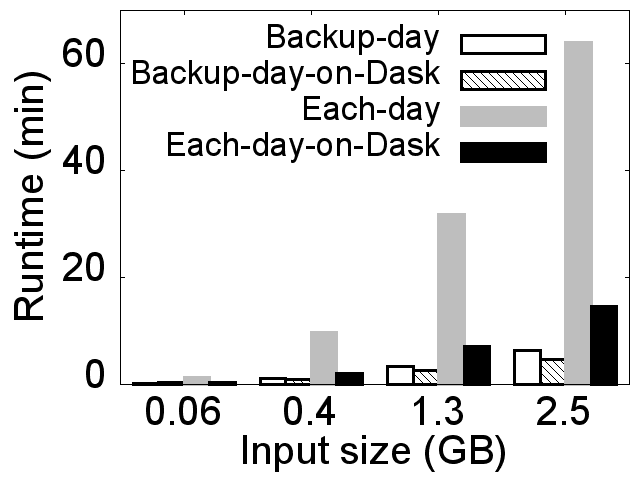}
    	\label{fig:runtime-dask}
	}
\caption{Runtime and scalability evaluation}
\label{fig:exp_runtime}
\end{figure}

%% file: sections/impact-evaluation.tex
\begin{figure}[!t]
\centering
	\subfigure[Backup scheduling]{
    	\includegraphics[width=0.3\columnwidth, keepaspectratio]{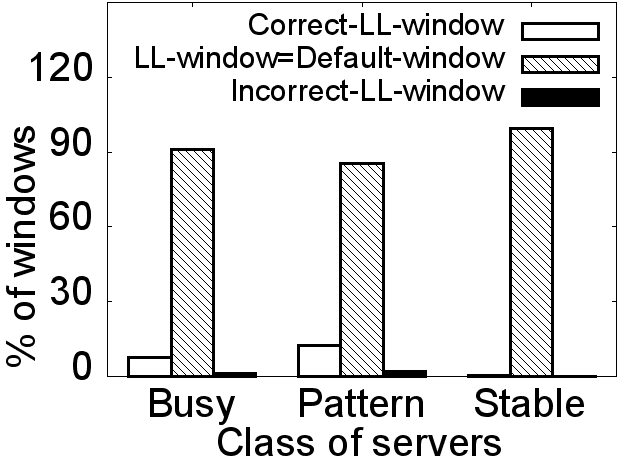}
    	\label{fig:backups}
	}\hspace{2mm}
	\subfigure[Auto-scale]{
    	\includegraphics[width=0.3\columnwidth, keepaspectratio]{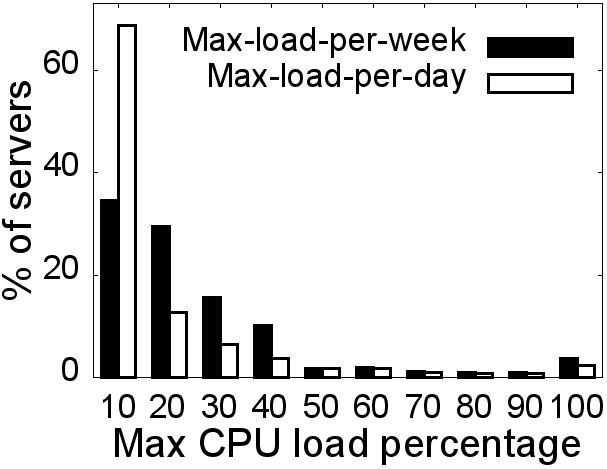}
    	\label{fig:autoscale}
	}
\caption{Impact evaluation per use case}
\label{fig:exp_impact}
\end{figure}

%% file: sections/lessons.tex
\section{Lessons Learned}
\label{sec:lessons}

\textbf{Keep Version One Simple}.
We originally started building the \app\ infrastructure to predict the load of several millions of SQL databases and enable preemptive auto-scale of resources. 
Since this was a complex and risky endeavor, we first tested the infrastructure on a \textit{smaller fleet} of tens of thousands of PostgreSQL and MySQL servers and a \textit{less risky scenario} of backup scheduling. 
We closely monitored, optimized, and adjusted the infrastructure since its deployment. Next, we will apply this matured system to more ambitious and risky scenarios and at higher scale.

\textbf{Verify Assumptions}.
Building a reliable and scalable infrastructure like \app\ is challenging. Thus, it is important to verify all assumptions that such a project relies upon. For example, when we started building \app, the mechanisms that scale resources of SQL databases were slow. Therefore, reactive auto-scale was unreliable and preemptive policies were needed. However, in the meantime, these mechanisms were optimized making even reactive auto-scale suitable for many databases. This example illustrates the need for a generic infrastructure to minimize loss of effort and amplify impact.

%% file: sections/related_work.tex
\section{Related Work}
\label{sec:related}

\textbf{Systems for ML} were proposed in the past. However, most of them lack easy integration with Azure compute~\cite{mlflow, tensorflow, velox, clipper, caffe, mllib}. In particular, security, privacy, license, compatibility, and interfaces would have to be done from scratch. 
Thus, we built the \app\ infrastructure upon Azure products~\cite{appinsights, data_lake, aml, cosmosdb}.
They offer all features we needed to build \app.
We also considered leveraging the model-serving infrastructure Resource Central~\cite{RC}. However, at that time Azure ML~\cite{aml} provided support and integration for a broader set of modeling and tracking tools.  



\textbf{Load Prediction} for optimized resource allocation on a cluster has become a popular research direction in the recent years. 
Existing approaches focus on
predicting survivability of databases for optimized resource provisioning~\cite{PLT18},
idle time detection for database quiescing and overbooking~\cite{LRDXGKC16, VCLRPKDH17}, 
database workload prediction for database consolidation~\cite{CJMB11},
VM workload prediction~\cite{KYTA12} for oversubscribing servers~\cite{RC}, dynamic VM provisioning~\cite{CMB14}, and reducing performance interference between VMs co-located on the same physical machine~\cite{deepdive},
workload classification for capacity planning and task scheduling~\cite{MHCD10},
cost- and QoS-aware application placement in virtualized server clusters~\cite{pMapper, bubble-flux},
and preemptive auto-scale of resources~\cite{DLNK16, quasar, dhalion, IKLL12, KRKM17, cloudscale, autocontrol, RDG11, pstore, press}. 
None of these approaches focused on predicting low load windows for optimized scheduling of system maintenance tasks. Thus, these approaches neither define low load prediction accuracy, nor compare ML models from the perspective of low load prediction. 

\textbf{Job Scheduling Algorithms} were proposed in the literature~\cite{rayon, morpheus, heracles}. Our backup scheduling algorithm (Section~\ref{sec:infrastructure}) is not the focus of this paper. It is just one example how the \app\ infrastructure can be used in production for optimized resource allocation on the cloud.


%% file: sections/conclusions.tex
\section{Conclusions}
\label{sec:conclusions}

We built the \app\ infrastructure for load prediction and optimized resource allocation on the cloud. While the infrastructure is applicable to a wide range of use cases, we illustrated it by the backup scheduling scenario.

%% file: sections/appendix.tex
\section{Preemptive Auto-scale of SQL Databases}
\label{sec:appendix}

As a follow-up project, we will use \app\ infrastructure (Figure~\ref{fig:infrastrucute}) for preemptive auto-scale of resources for Azure SQL databases (Example~2 in Section~\ref{sec:introduction}). Below, we briefly summarize our initial results in database classification and load prediction. 

\subsection{Classification of SQL Databases}
\label{sec:sql-data}

SQL data contains database identifier, timestamp in minutes, and average CPU load per 15 minutes. We differentiate between stable and unstable databases.

\begin{definition}(\textbf{Stable Database})
A stable database is defined as a database whose variation does not exceed one standard deviation for the last three days in the period evaluated. 
Otherwise, a database is called unstable.
\label{def:stable-db}
\end{definition}

We analyzed a random sample of several thousands of single standard and premium SQL databases during one month in 2019 and concluded that 19.36\% of them are stable.

\subsection{Prediction Error Metrics}
\label{sec:sql-metrics}

For the preemptive auto-scale use case, we predict the CPU load per database 24 hours ahead. 
We define error as the difference between the forecast and the true load in Equation~1.
We use the standard metrics, namely, Mean Normalized Root Mean Squared Error (Mean NRMSE) and Mean Absolute Scaled Error (MASE) to evaluate accuracy of models in Equations~2 and 3. 

\begin{align}
\mathit{error} &= \mathit{forecast} - \mathit{true} \\  
\mathit{Mean\ NRMSE} &= \frac{\sqrt{mean(error^2)}}{mean(true)} \\
\mathit{MASE} &= \mathit{mean}\left(\frac{abs.(error)}{\mathit{normalizing\ factor}}\right) 
\end{align}

\begin{figure}[!t]
\centering
\includegraphics[width=.8\columnwidth]{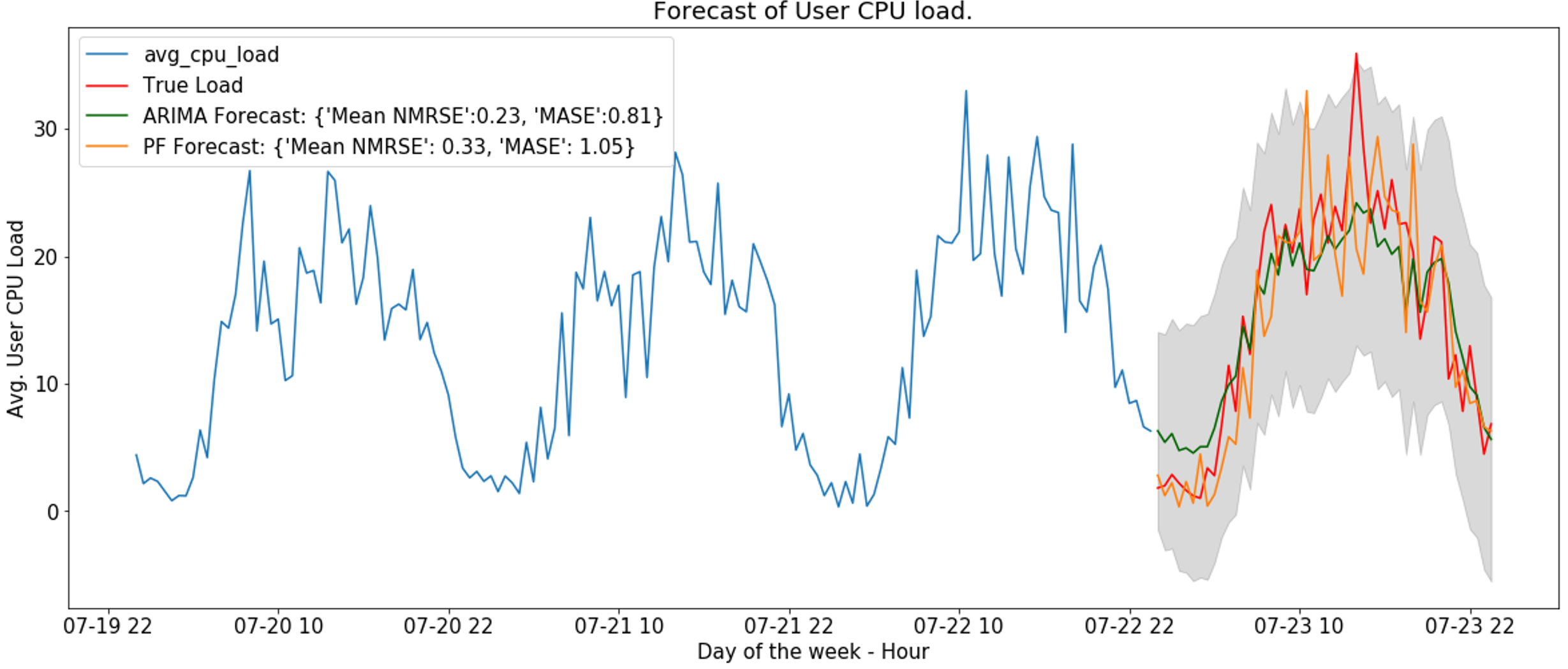}
\caption{Accurate load prediction}
\label{fig:sql-error-good}
\end{figure}

\begin{figure}[!t]
\centering
\includegraphics[width=.8\columnwidth]{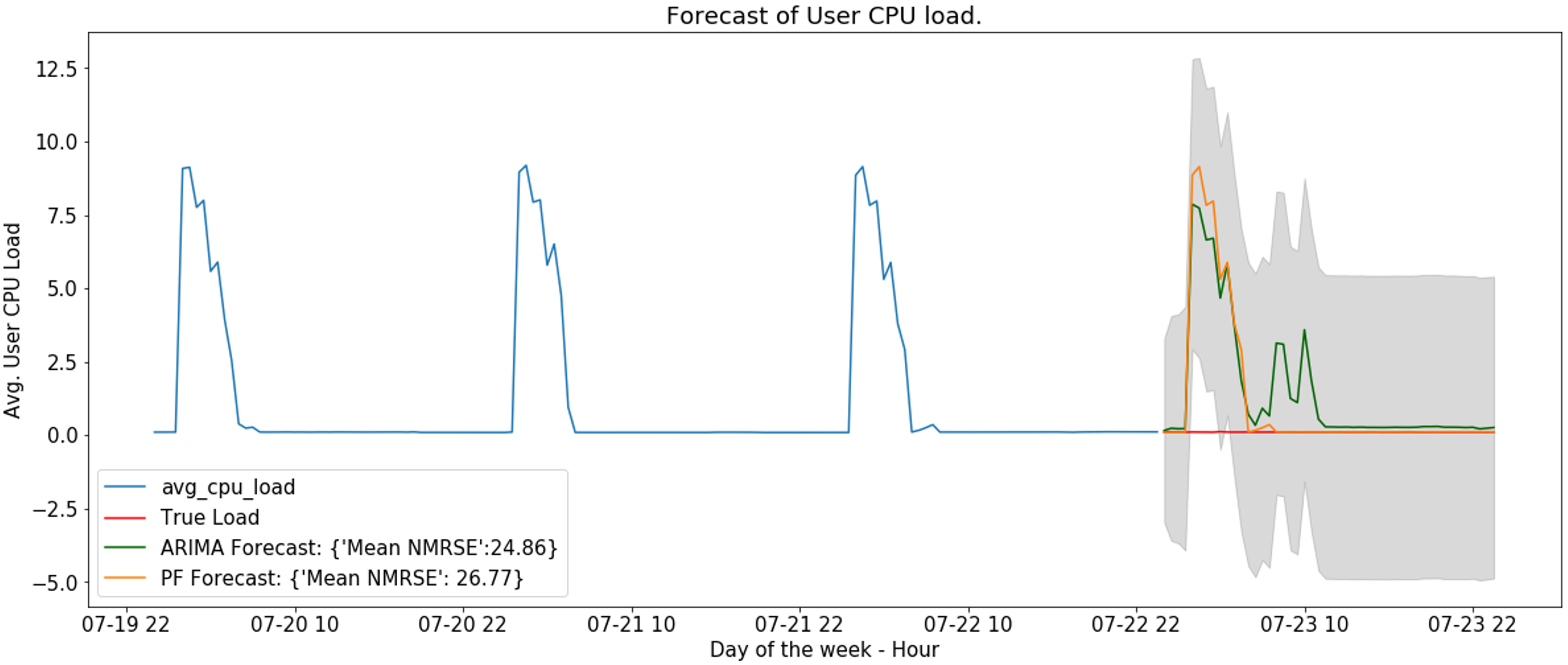}
\caption{Inaccurate load prediction}
\label{fig:sql-error-bad}
\end{figure}

A mean NRMSE of 1 is produced when the mean is predicted as the forecast, anything less than 1 would mean doing better than forecasting the mean. 
The normalizing factor in this case is the error produced by a one step ahead true forecast. MASE of less than 1 means doing better than a one step ahead true forecast. 
Figures~\ref{fig:sql-error-good} and \ref{fig:sql-error-bad} show examples of values of these metrics. 

\subsection{Load Prediction}
\label{sec:sql-predictions}

\begin{figure}[!t]
\centering
\includegraphics[width=.5\columnwidth]{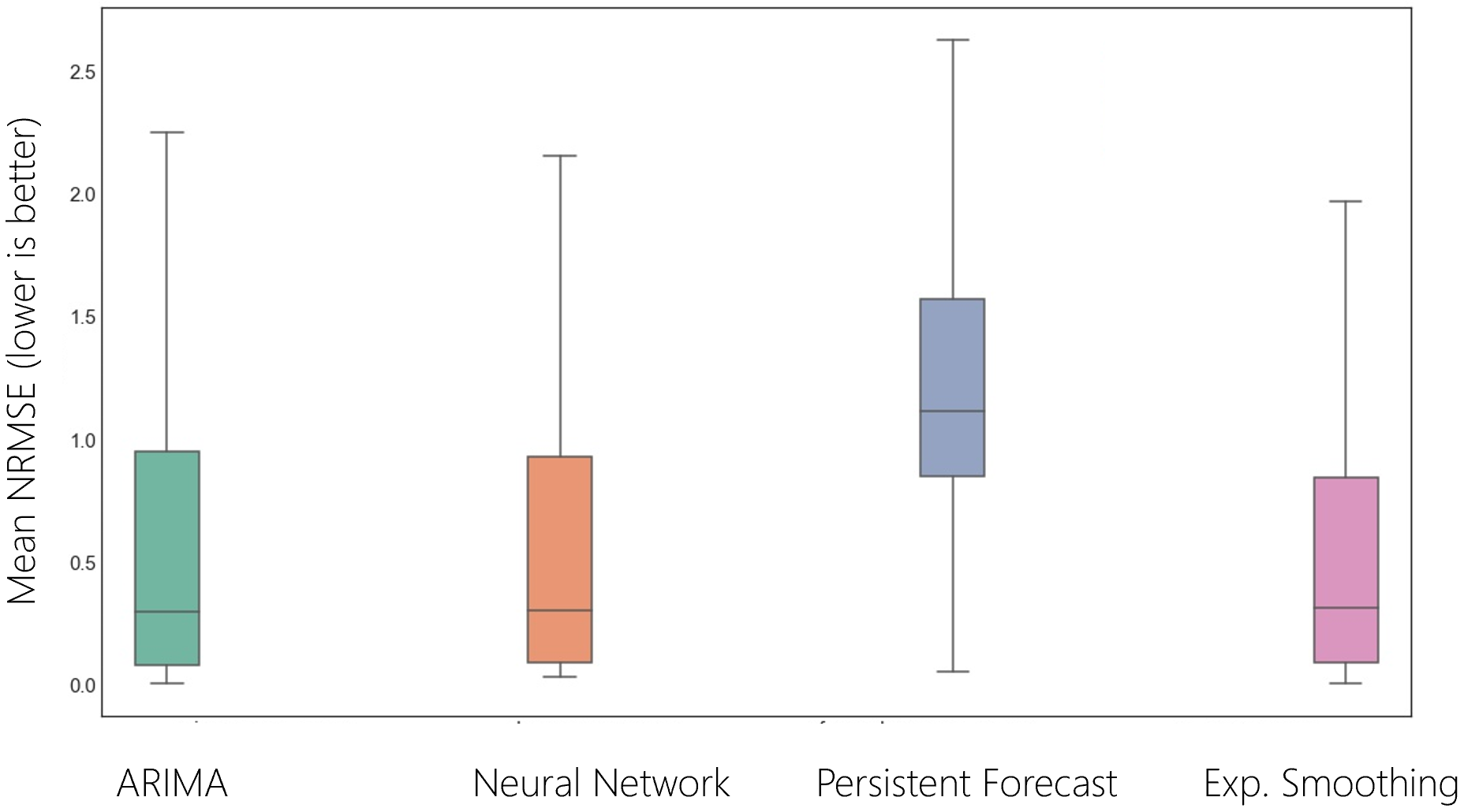}
\caption{Model accuracy}
\label{fig:sql-accuracy}
\end{figure}

\begin{figure}[!t]
\centering
\includegraphics[width=.55\columnwidth]{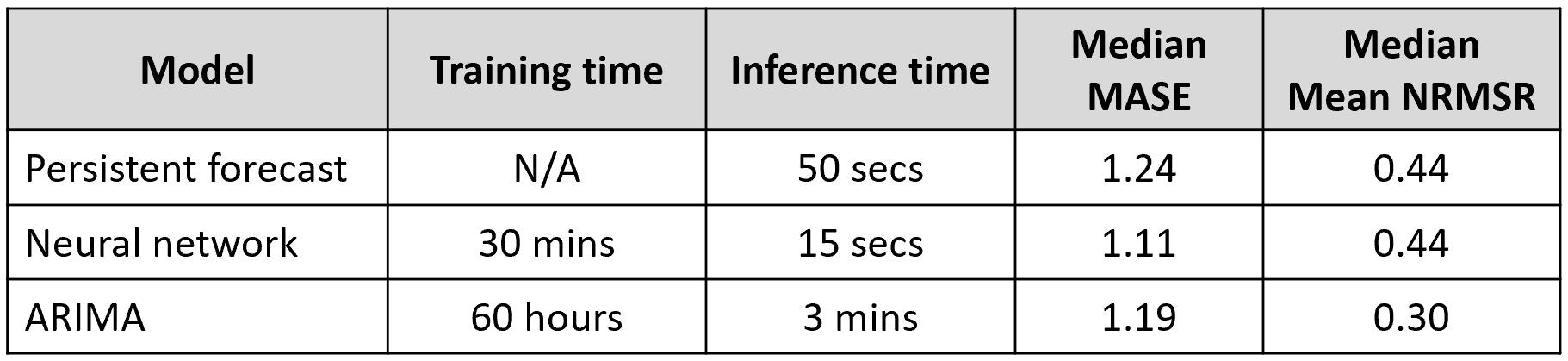}
\caption{Training, inference, and accuracy}
\label{fig:sql-accuracy-2}
\end{figure}

Figures~\ref{fig:sql-accuracy} and \ref{fig:sql-accuracy-2} summarize the accuracy and runtime of training and inference per ML model described in  Section~\ref{sec:models}. 
Neural network refers to GluonTS~\cite{gluonts} and persistent forecast is based on previous day.
GluonTS and ARIMA are trained on one week of historical load per database.
ARIMA runs in parallel per database on HDI cluster with 2 head nodes  with 4 cores and 28GB of memory per node and 2 worker nodes with 4 cores and 56GB per node. Given the coarse granularity of SQL data (per 15 minutes), ARIMA works better than for the fine-grained PostgreSQL and MySQL data (per 5 minutes). Nevertheless, the runtime for training of ARIMA is still not comparable with other models.
Based on this preliminary evaluation, we concluded that for SQL databases persistent forecast also finds the middle ground between accuracy and computational overhead.

